\documentclass[
6x9    ,final            
  ]
  {aipproc}

\layoutstyle{6x9}

\begin{document}

\title{Overview of Issues Surrounding Strangeness in the Nucleon }


\author{Anthony W. Thomas}{
  address={Suite 1, Jefferson Lab,12000 Jefferson Ave., Newport News VA 23606 USA \\ and College of William and Mary, Williamsburg VA 23187 USA}
}

\begin{abstract}
The calculation of the strangeness content of the nucleon and its 
experimental verification is a fundamental step in establishing 
non-perturbative QCD as the correct theory describing the 
structure of hadrons. It holds a role in QCD analogous to the 
correct calculation of the Lamb shift in QED. We review the 
latest developments in the vector and scalar matrix elements of 
the strange quarks in the proton, where there has recently been 
considerable progress.
\end{abstract}

\maketitle


\section{Introduction}
Over the past decade there have been heroic efforts at MIT-Bates, Mainz and 
JLab~\cite{Armstrong:2005hs,Aniol:2005zf,Maas:2004dh,Hasty:2001ep} 
to use parity violating electron scattering to extract the vector matrix 
elements of the strange quarks in the proton, 
$\langle \bar{s} \gamma^\mu s \rangle$. These measurements have allowed a 
careful global analysis from which it has been established that less than 5\% of 
the magnetic moment and less than 2\% of the charge radius of the proton can 
be attributed to strange quarks~\cite{Young:2006jc,Young:2007zs}. 
At this meeting the G0 Collaboration reported 
preliminary results from its final backward angle run~\cite{G0_CIPANP}, 
in which new information 
on the strange magnet form factor and the axial form factor at high 
$Q^2$ were presented.  On the theoretical side there was also a presentation from the University of 
Kentucky group concerning the direct measurement of the strange magnetic 
form factor~\cite{KFL}. 

The scalar form factors of the nucleon have been of great theoretical interest for 
decades, because the $\pi$ N and strange quark sigma commutators
\begin{equation}
\sigma_{\pi N} \, = \, \frac{(m_u \, + \, m_d)}{2} \,  
\langle \bar{u} u \, + \, \bar{d} d \rangle \, \, ; \, \,
\sigma_s \, = \, m_s \, \langle \, \bar{s} s \rangle \,  ,
\label{eq:sigma}
\end{equation}
are directly related to the chiral symmetry breaking in the QCD Hamiltonian. 
Both have been somewhat controversial in terms of their extraction from 
experimental data and in fact it seems unlikely that the strange sigma 
commutator, which has long been believed to account for as much as one 
third of the mass of the nucleon,  
will ever be extracted from data with any degree of precision. Currently, these 
terms are of some practical importance in an unexpected quarter, namely the 
search for dark matter. Within the constrained, minimal supersymmetric extensions 
of the Standard Model, the neutralinos are a promising candidate for dark matter 
and their interaction with hadronic matter is determined by the 
sigma commutators~\cite{Ellis:2008hf}. 
This means that the interpretation of the results of dark matter searches depends 
strongly on how accurately one can determine $\sigma_{\pi N}$ and $\sigma_s$. 
As we shall explain, the answer to this need has come from an unexpected source, 
namely the  study of hadron properties as a function of quark mass using 
lattice QCD~\cite{Young:2009zb}.

\section{Vector Strange Form Factors}
The determination of the contribution to nucleon properties from quark loops, 
the so-called ``disconnected terms'' in lattice QCD, has proven very difficult 
by direct means. This led to the formulation of indirect 
methods~\cite{Leinweber:1999nf}, which have proven 
extremely effective in extracting accurate values for the strange contributions to 
the electric and magnetic form factors~\cite{Leinweber:2004tc,Leinweber:2006ug}. 
For example, for the strangeness magnetic 
moment was determined by these means to be 
$-0.046 \pm 0.019 \, \mu_N$~\cite{Leinweber:2004tc}, which 
at just a few hundredths of a nuclear magneton represents a remarkably accurate 
determination. 

Recently, there has been significant progress in the development  of direct methods 
for calculating the contribution from quark loops. First, for the measurement of 
the moments of parton distribution functions, the University of Kentucky group 
employed sophisticated numerical methods to extract a non-zero signal for 
the momentum fraction carried by strange and anti-strange quarks in the 
proton~\cite{Deka:2008xr}, namely 
$\langle x(s \, + \, \bar{s} ) \rangle = 0.027 \pm 0.006$ -- albeit at 
a somewhat large light quark mass. This was recently followed~\cite{Doi:2009sq}, 
as reported at this 
meeting by Keh-Fei Liu~\cite{KFL}, by a clear non-zero signal at a range of 
momentum transfer values for 
the strangeness magnetic form factor. At the large light quark masses employed, 
the value obtained at zero momentum transfer was 
$G_M^s(Q^2=0) = - 0.017 
\pm 0.025 \pm 0.07 \mu_N$. 
Using the dependence on light quark mass 
found in Ref.~\cite{Wang:2007iw}, this would be expected to increase 
in magnitude by about 80\% 
at the physical light quark mass. With or without the latter correction, this 
direct determination is clearly in excellent quantitative agreement with the 
earlier calculation of Leinweber {\it et al.}~~\cite{Leinweber:2004tc}.

The best experimental determination of the strange 
magnetic form factor, at $Q^2 
= 0.1$ GeV$^2$, from a global analysis of 
all published data~\cite{Young:2007zs}, is 
$-0.01 \pm 0.25 \mu_N$. Clearly this is in 
very good agreement with the theoretical 
values. However, in a unique example for 
strong interaction physics, the theoretical 
calculations are an order of magnitude more precise than the state of the art 
experiments! This makes the quest for really bright new ideas to improve the 
experimental accuracy very important indeed.

With the theoretical and experimental values of the strange form factors pinned 
down near $Q^2 = 0$, it is interesting to also explore the dependence on 
$Q^2$. At higher $Q^2=0.22$ GeV$^2$, the A4 Collaboration at 
Mainz recently reported a new value 
of the strange magnetic form factor~\cite{Baunack:2009gy}, 
namely $-0.14 \pm 0.11 \pm 0.11 \mu_N$, 
again in very good agreement with the latest 
application of the indirect methods~\cite{Wang:1900ta} 
$G_M^s(Q^2=0.22) = - 0.034 \pm 0.031 \mu_N$. 
The G0 Collaboration reported 
a preliminary analysis of its back angle run at this conference, with the value 
at 0.23 GeV$^2$ consistent with the Mainz measurement~\cite{G0_CIPANP}. 
It also seems likely that 
the collaboration will determine the axial form factor at the larger $Q^2$ 
values.

\subsection{Scalar Form Factor}
The strange sigma commutator, $\sigma_s$, is tricky to measure directly 
in lattice QCD because it involves a subtraction of the strange quark loop 
in vacuum from that in the nucleon and as we shall see the difference is 
relatively small. While the common belief is that it is of order 1/3 of the 
mass of the nucleon~\cite{Nelson:1987dg}, 
the first hint that it may be much smaller came in a study 
made in connection with the possible experimental determination of a time 
dependent variation in the fundamental ``constants'' 
of Nature~\cite{Flambaum:2004tm}. However, it 
is only this year, with the analysis of a several independent, high precision data sets 
on the masses of the nucleon octet, 
within full 2+1 flavor QCD~\cite{Aoki:2008sm,Lin:2008pr}, 
that it has been possible to make a precise mass formula which incorporates the 
correct non-analytic behavior and reproduces all of the data in a convincing 
manner~\cite{Young:2009zb}. 
The mass formula involves the usual SU(3) expansion to first order in the 
quark masses, plus the one loop chiral corrections including the Goldstone 
boson masses and evaluated using finite range regularization.

This procedure not only produces an 
excellent fit to all of the data but the octet 
masses extrapolated to the physical point 
all agree with experiment at the 2\% level 
or better. Given the expressions for the masses versus $m_\pi$ and $m_K$ one 
can directly evaluate the sigma terms using the Feynman-Hellman theorem. 
The result for $\sigma_{\pi N} \, = \, 47 \pm 10$ MeV is certainly consistent 
with most studies. However, the result for the strange term, $\sigma_s \, 
= \, 31 \pm 16$ MeV is an order of magnitude smaller than the classic result. 
It is this order of magnitude 
reduction that is expected to have profound implications 
for searches for dark matter~\cite{inprogress}. 

After the determination by Young and Thomas~\cite{Young:2009zb}, 
a new direct calculation of 
$\sigma_s$ was reported by Touassaint and Freeman~\cite{Toussaint:2009pz}. 
Their value of 
$59 \pm 11 $ MeV is consistent with that reported above and appears to 
confirm that $\sigma_s$ is considerably smaller than hitherto believed. 

\section{Conclusion}
The last few years have seen remarkable progress in both the theoretical 
and experimental determination of the strange quark matrix elements in 
the proton. For the present time the theoretical calculations hold the precision 
lead with a great need for new ideas if the experimental determinations are 
to reach a similar level. Nevertheless, within the currently possible limits, 
QCD works very well and by analogy with the Lamb 
shift in QED, non-perturbative QCD has satisfied a crucial test.


\begin{theacknowledgments}
This work was supported by U.S. DOE Contract No. DE-AC05-06OR23177, 
under which Jefferson Science Associates, LLC operates Jefferson 
Laboratory.
\end{theacknowledgments}



\begin{thebibliography}{9}
%
\bibitem{Armstrong:2005hs}
  D.~S.~Armstrong {\it et al.}  [G0 Collaboration],
  Phys.\ Rev.\ Lett.\  {\bf 95}, 092001 (2005)
  [arXiv:nucl-ex/0506021].
%
\bibitem{Aniol:2005zf}
  K.~A.~Aniol {\it et al.}  [HAPPEX Collaboration],
  Phys.\ Rev.\ Lett.\  {\bf 96}, 022003 (2006)
  [arXiv:nucl-ex/0506010].
%
\bibitem{Maas:2004dh}
  F.~E.~Maas {\it et al.},
  Phys.\ Rev.\ Lett.\  {\bf 94}, 152001 (2005)
  [arXiv:nucl-ex/0412030].
%
\bibitem{Hasty:2001ep}
  R.~Hasty {\it et al.}  [SAMPLE Collaboration],
  Science {\bf 290}, 2117 (2000)
  [arXiv:nucl-ex/0102001].
%
%
\bibitem{Young:2006jc}
  R.~D.~Young, J.~Roche, R.~D.~Carlini and A.~W.~Thomas,
  Phys.\ Rev.\ Lett.\  {\bf 97}, 102002 (2006)
  [arXiv:nucl-ex/0604010].
%
\bibitem{Young:2007zs}
  R.~D.~Young, R.~D.~Carlini, A.~W.~Thomas and J.~Roche,
  Phys.\ Rev.\ Lett.\  {\bf 99}, 122003 (2007)
  [arXiv:0704.2618 [hep-ph]].
%
\bibitem{G0_CIPANP}
B. Beise [G0 Collaboration], invited talk at CIPANP (May 2009)
%
\bibitem{KFL}
K.-F. Liu, invited talk at CIPANP (May 2009)
%
\bibitem{Ellis:2008hf}
  J.~R.~Ellis, K.~A.~Olive and C.~Savage,
  Phys.\ Rev.\  D {\bf 77}, 065026 (2008)
  [arXiv:0801.3656 [hep-ph]].
%
\bibitem{Young:2009zb}
  R.~D.~Young and A.~W.~Thomas,
  arXiv:0901.3310 [hep-lat].
%
\bibitem{Leinweber:1999nf}
  D.~B.~Leinweber and A.~W.~Thomas,
  Phys.\ Rev.\  D {\bf 62}, 074505 (2000)
  [arXiv:hep-lat/9912052].
%
\bibitem{Leinweber:2004tc}
  D.~B.~Leinweber {\it et al.},
  Phys.\ Rev.\ Lett.\  {\bf 94}, 212001 (2005)
  [arXiv:hep-lat/0406002].
%
\bibitem{Leinweber:2006ug}
  D.~B.~Leinweber {\it et al.},
  Phys.\ Rev.\ Lett.\  {\bf 97}, 022001 (2006)
  [arXiv:hep-lat/0601025].
%
\bibitem{Deka:2008xr}
  M.~Deka {\it et al.},
  Phys.\ Rev.\  D {\bf 79}, 094502 (2009)
  [arXiv:0811.1779 [hep-ph]].
%
\bibitem{Doi:2009sq}
  T.~Doi {\it et al.},
  arXiv:0903.3232 [hep-ph].
%
\bibitem{Wang:2007iw}
  P.~Wang, D.~B.~Leinweber, A.~W.~Thomas and R.~D.~Young,
  Phys.\ Rev.\  D {\bf 75}, 073012 (2007)
  [arXiv:hep-ph/0701082].
%
\bibitem{Baunack:2009gy}
  S.~Baunack {\it et al.},
  Phys.\ Rev.\ Lett.\  {\bf 102}, 151803 (2009)
  [arXiv:0903.2733 [nucl-ex]].
%
\bibitem{Wang:1900ta}
  P.~Wang, D.~B.~Leinweber, A.~W.~Thomas and R.~D.~Young,
  arXiv:0807.0944 [hep-ph].
%
\bibitem{Nelson:1987dg}
  A.~E.~Nelson and D.~B.~Kaplan,
  Phys.\ Lett.\  B {\bf 192}, 193 (1987).
%
\bibitem{Flambaum:2004tm}
  V.~V.~Flambaum, D.~B.~Leinweber, A.~W.~Thomas and R.~D.~Young,
  Phys.\ Rev.\  D {\bf 69}, 115006 (2004)
  [arXiv:hep-ph/0402098].
%
\bibitem{Aoki:2008sm}
  S.~Aoki {\it et al.}  [PACS-CS Collaboration],
  arXiv:0807.1661 [hep-lat].
%
\bibitem{Lin:2008pr}
  H.~W.~Lin {\it et al.}  [Hadron Spectrum Collaboration],
  Phys.\ Rev.\  D {\bf 79}, 034502 (2009)
  [arXiv:0810.3588 [hep-lat]].
%
\bibitem{inprogress}
J.~Giedt, A.~W.~Thomas and R.~D.~Young, to be published.
%
\bibitem{Toussaint:2009pz}
  D.~Toussaint and W.~Freeman  [MILC Collaboration],
  arXiv:0905.2432 [hep-lat].
%
\end{thebibliography}
\end{document}